**Aleksandr CARIOW[1], Galina CARIOWA[1]**

[1]WEST POMERANIAN UNIVERSITY OF TECHNOLOGY, SZCZECIN, Żołnierska St. 49, 71-210, Szczecin Poland

# Hardware-Efficient Structure of the Accelerating Module for Implementation of Convolutional Neural Network Basic Operation

**Abstract**

This paper presents a structural design of the hardware-efficient module for implementation of convolution neural network (CNN) basic operation with reduced implementation complexity. For this purpose we utilize some modification of the Winograd's minimal filtering method as well as computation vectorization principles. This module calculate inner products of two consecutive segments of the original data sequence, formed by a sliding window of length 3, with the elements of a filter impulse response. The fully parallel structure of the module for calculating these two inner products, based on the implementation of a naïve method of calculation, requires 6 binary multipliers and 4 binary adders. The use of the Winograd's minimal filtering method allows to construct a module structure that requires only 4 binary multipliers and 8 binary adders. Since a high-performance convolutional neural network can contain tens or even hundreds of such modules, such a reduction can have a significant effect.

## 1. Introduction

Artificial intelligence, deep learning, and neural networks represent powerful and incredibly effective machine learning-based techniques used to solve many scientific and practical problems. Applications of deep neural networks to machine learning are diverse and promptly developing, reaching the various fields of fundamental sciences, technologies and real-world. Among the various types of deep neural networks, convolutional neural networks (CNN) are the most widely used [1]. The basic and most time-consuming operation in CNN is the operation of a two-dimensional convolution. Several methods have been proposed to accelerate the calculation of convolution, including the reduction of arithmetic operations via Fast Fourier transform (FFT) and the use of hardware accelerators based on FPGA, GPU and ASIC [2-16]. FFT based method of computing convolution is traditionally used for large filters, but state of the art CNN use small filters. In this situation one of the most effective algorithms used in the computation of a small-length two-dimensional convolution is the Winograd's minimal filtering algorithm, that is most intensively used in recent time [17]. The algorithm compute linear convolution over small tiles with minimal complexity, which makes it more effective with small filters and small batch sizes. In fact, this algorithm calculates two inner products of neighboring vectors formed by a sliding time window from the current data stream with an impulse response of the 3-tap finite impulse response (FIR) filter.

Many publications have been devoted to the implementation of computations in networks based on the Winograd's minimal filtering method [17-20]. However, the principles of organizing the structure of the module that implements the filtering algorithm have not been considered in detail by anyone. Our publication is intended to fill this gap.

## 2. Preliminaries

As already noted, the basic operation of convolutional neural networks is a sliding inner product of vectors, formed by a moving time window from the current data stream with an impulse response of the $M$-tap FIR filter. It can be described by the following formula:

$$y_l = \sum_{i=0}^{M-1} x_{i+l} h_i, \quad i = 0,1,...,M-1, \quad l = 0,1,...,N-M+1,$$

where $x_i$ are the elements of the current data stream, $h_i$ are the elements of the impulse response of FIR filter, which are constants.

Direct calculation of the inner product of two vectors of length $M$ requires $M$ multiplication and $M$-1 additions.

Direct application of two consecutive steps of a 3-tap FIR filter with coefficients $\{h_0, h_1, h_2\}$ to a set of 4 elements $\{x_0, x_1, x_2, x_3\}$ requires 6 additions and 6 multiplications:

$$y_0 = x_0 h_0 + x_1 h_1 + x_2 h_2, \quad y_1 = x_1 h_0 + x_2 h_1 + x_3 h_2 \quad (1)$$

Fig. 1 explains the essence of the reasoning.

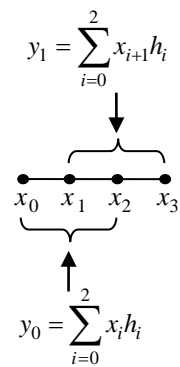

Fig. 1.　Demonstration of the essence of computation execution in accordance with the Winograd's minimal filtering method

The idea of Winograd's minimal filtering method is to compute these two filter outputs in following way [17]:

$$\mu_1 = (x_0 - x_2)h_0, \quad \mu_2 = (x_1 + x_2)\frac{h_0 + h_1 + h_2}{2},$$

$$\mu_3 = (x_2 - x_1)\frac{h_0 - h_1 + h_2}{2}, \quad \mu_4 = (x_1 - x_2)h_2,$$

$$y_0 = \mu_1 + \mu_2 + \mu_3, \quad y_1 = \mu_2 - \mu_3 - \mu_4.$$

The values $(h_0 + h_1 + h_2)/2$ and $(h_0 - h_1 + h_2)/2$ can be calculated in advance, then this method requires 4 multiplications and 8 additions, which is equal to number of arithmetical operations in the direct method. But since multiplication is a much more complicated operation than addition, the Winograd's minimal filtering method is more efficient than the direct method of computation.

The above expressions exhaustively describe the entire set of mathematical operations needed to compute, but they do not disclose the way and sequence of the computation organization, nor the structure of the processor module that implements these operations.

## 3. Structural synthesis of Winograd's minimal filtering module

Let $\mathbf{X}_{4\times 1} = [x_0, x_1, x_2, x_3]^T$ - be a column vector, that represent the input tile, $\mathbf{H}_{3\times 1} = [h_0, h_1, h_2]^T$ - be a column vector, that contains the all coefficients of impulse response of 3-tap FIR filter (filter tile), and $\mathbf{Y}_{4\times 1} = [y_0, y_1]^T$ - be a column vector containing result of computing two outputs the 3-tap FIR filter. Then, a fully parallel algorithm for computation $\mathbf{Y}_{4\times 1}$ using Winograd's minimal filtering method can be written with the help of following matrix-vector calculating procedure:

$$\mathbf{Y}_{4\times 1} = \mathbf{A}^{(2)}_{2\times 4} diag(\mathbf{S}_{4\times 1}) \mathbf{A}^{(1)}_{4} \mathbf{X}_{4\times 1} \quad (2)$$

where

$$\mathbf{A}^{(1)}_{4} = \begin{bmatrix} 1 & -1 \\ 1 & 1 \\ -1 & 1 \\ 1 & -1 \end{bmatrix}, \quad \mathbf{A}^{(2)}_{2\times 4} = \begin{bmatrix} 1 & 1 & 1 & 0 \\ 0 & 1 & -1 & -1 \end{bmatrix},$$

and $\mathbf{S}_{4\times 1} = [s_0, s_1, s_2, s_3]^T$, $s_0 = h_0$, $s_1 = (h_0 + h_1 + h_2)/2$, $s_2 = (h_0 - h_1 + h_2)/2$, $s_3 = h_2$.

Entries of the matrix $diag(\mathbf{S}_{4\times 1})$ can be computed with the help of the following procedure:

$$\mathbf{S}_{4\times 1} = \mathbf{D}_4 \breve{\mathbf{A}}^{(2)}_{4\times 5} \breve{\mathbf{A}}^{(1)}_{5\times 3} \mathbf{H}_{3\times 1} \quad (3)$$

where

$$\mathbf{D}_4 = diag(1, 1/2, 1/2, 1),$$

$$\breve{\mathbf{A}}^{(1)}_{5\times 3} = \begin{bmatrix} 1 & & \\ & 1 & \\ 1 & & 1 \\ & -1 & \\ & & 1 \end{bmatrix}, \quad \breve{\mathbf{A}}^{(2)}_{4\times 5} = \begin{bmatrix} 1 & & & & \\ & 1 & 1 & & \\ & & 1 & 1 & \\ & & & & 1 \end{bmatrix},$$

Fig. 2 shows a data flow diagram of the proposed algorithm for the implementation of Winograd's minimal filtering basic operation. In this paper, data flow diagrams are oriented from left to right and Fig. 3 shows a data flow diagram of the process for calculating the vector $\mathbf{S}_{4\times 1}$ entries. Straight lines in the figures denote the operations of data transfer. The circles in these figures show the operation of multiplication by a real number inscribed inside a circle. Points where lines converge denote summation a dotted lines indicate the sign-change operations. We use the usual lines without arrows on purpose, so as not to clutter the picture.

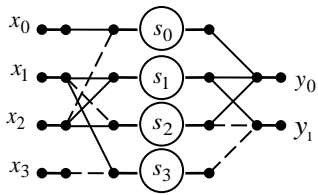

Fig. 2. The data flow diagram of proposed algorithm for implementation of Winograd's minimal filtering basic operation

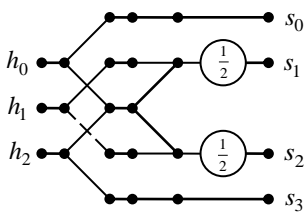

Fig. 3. The data flow diagram describing the process of calculating entries of the vector $\mathbf{S}_{4\times 1}$ in accordance with the procedure (3).

In low power application specific integrated circuits (ASIC) design, optimization must be primarily done at the level of transistor amount. From this point of view a multiplication requires much more intensive hardware resources than an addition. Moreover, a binary multiplier occupies much more area and consumes much more power than binary adder. This is because the implementation complexity of a fully parallel multiplier grows quadratically with operand length, while the implementation complexity of an adder increases linearly with operand length. Therefore, the algorithm containing as little as possible of multiplications is preferable from the point of view of ASIC design.

Fig. 4 shows a structure of processing module for ASIC-oriented implementation of Winograd's minimal filtering basic operation. The module contains four two-input and two three-input algebraic adders, four multipliers and a register memory for storing the values $s_i$. It is assumed that these elements can be precomputed and written to the register memory before the calculations begin. Depending on the requirements for the speed of calculations, the modules can be cascaded and combined into clusters.

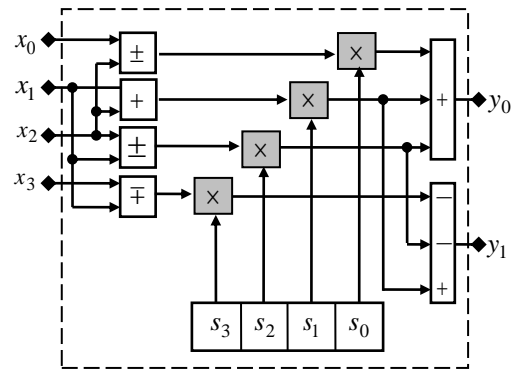

Fig. 4. The structure of the processor module for implementing the Winograd's minimum filtering operation (ASIC point of view).

Today a better alternative than ASIC are FPGAs (field-programmable gate arrays) - the integrated circuits designed to be configured by a customer or a designer. If the early FPGAs contained only small embedded multipliers, then more recent FPGAs contain DSP blocks, that include not only multipliers, but also internal adders designed in such a way that part of the additions in (1) can also be computed inside the DSP blocks. However even if the DSP block contains embedded multipliers, their number is always limited. This means that if the implemented scheme has a large number of multiplications, the projected processor may not always fit into the chip and the problem of minimizing the number of multipliers remains relevant.

This applies fully to FPGAs Stratix II that contain DSP blocks, each of which includes just 4 multipliers, as well as three adders at the block input and three adders at the block output. Such a block structure allows using the hardware resources of the chip with a maximum degree of efficiency. It is easy to see that a fully parallel implementation of direct calculations does not fit into the boundaries of one Stratix II DSP block.

Fig. 5 shows a structure of processing module for implementation of Winograd's minimal filtering basic operation on the base of Altera Stratix II high-speed FPGA chip. The bulk of the computation is performed inside the DSP block, but the adders outlined by the dash-dotted line on a gray background are implemented using external logic gates. By way of background information, it is necessary to emphasize, that not all outputs of the DSP block are used in the proposed solution (see Fig. 5). Depending on the performance that is needed in the neural network, the number of processor modules implementing the Winograd's filtering basic operation can be quite large.

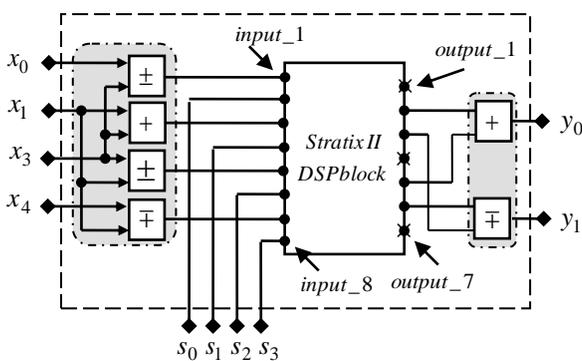

Fig. 5. The structure of the accelerating module for implementing the Winograd's minimum filtering basic operation (FPGA point of view).

## 4. Conclusion

This work looks into some issues of structural design of the hardware-efficient module for implementation of CNN basic operation using Winograd's minimal filtering method. This method reduces the number of multipliers at the expense of increased number of adders. Taking into account a relative hardware complexity of multiplier and adder, reducing the number of multipliers at the expense of the increased number of adders is desirable. The calculations demonstrate the effectiveness of proposed solutions and their universal impact on the different types of CNN layers as well as on the principles of network operation.

**Prof., DSc., PhD Aleksandr CARIOW**

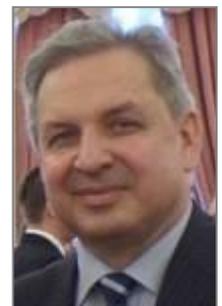

He received the Candidate of Sciences (PhD) and Doctor of Sciences degree (DSc or Habilitation) in Computer Sciences from LITMO of St. Petersburg, Russia in 1984 and 2001, respectively. In September 1999, he joined the faculty of Computer Sciences and Information Technology at the West Pomeranian University of Technology, Szczecin, Poland, where he is currently working as a professor in the Department of Computer Architectures and Teleinformatics. His research interests include digital signal processing algorithms, VLSI architectures, and data processing parallelization.

*e-mail: acariow@wi.zut.edu.pl*

**PhD Galina CARIOWA**

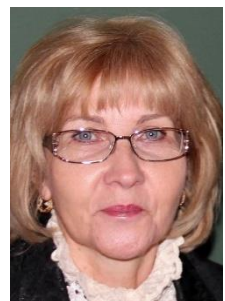

She received the MSc degree in Mathematics from Moldavian State University, Chişinău in 1978 and PhD degree in computer science from West Pomeranian University of Technology, Szczecin, Poland in 2007. She is currently working as an assistant professor of the Department of Computer Architectures and Teleinformatics. She is also an Associate-Editor of World Research Journal of Transactions on Algorithms. Her scientific interests include numerical linear algebra and digital signal processing algorithms, VLSI architectures, and data processing parallelization.

*e-mail: gcariowa@wi.zut.edu.pl*